\documentclass[10pt, conference, compsocconf]{IEEEtran}
\usepackage[pdftex]{graphicx}
\usepackage{xspace}

\begin{document}

\title{Data Intensive High Energy Physics Analysis in a Distributed Cloud}
\author{\IEEEauthorblockN{R.J. Sobie,
A. Agarwal,
M. Anderson,
P. Armstrong,
K. Fransham,
I. Gable,
D. Harris \\
C. Leavett-Brown,
M. Paterson,
D. Penfold-Brown,
M. Vliet}
\IEEEauthorblockA{Department of Physics and Astronomy \\
University of Victoria, Victoria, Canada V8W 3P6 \\
Email: rsobie@uvic.ca}
\vspace{0.5cm}
\IEEEauthorblockN{A. Charbonneau, R. Impey, W. Podaima}
\IEEEauthorblockA{National Research Council Canada \\
100 Sussex Drive, Ottawa, Canada}
}
\maketitle

\begin{abstract}
We show that distributed Infrastructure-as-a-Service (IaaS)
compute clouds can be effectively used for the analysis of high 
energy physics data. 
We have designed a distributed cloud system that works with any 
application using large input data sets requiring a high 
throughput computing environment.
The system uses IaaS-enabled science and commercial clusters 
in Canada and the United States.
We describe the process in which a user prepares an analysis virtual machine (VM)
and submits batch jobs to a central scheduler.
The system boots the user-specific VM on one of the IaaS clouds, runs the jobs 
and returns the output to the user.
The user application accesses a central database for calibration data during
the execution of the application.
Similarly, the data is located in a central location and streamed by the running
application.
The system can easily run one hundred simultaneous jobs in an efficient 
manner and should scale to many hundreds and possibly thousands of user jobs.
\end{abstract}

\IEEEpeerreviewmaketitle

\section{Introduction}

Infrastructure as a Service (IaaS) cloud computing is emerging as a new 
and efficient way to provide computing to the research community.
The growing interest in clouds can be attributed, in part, to the ease of 
encapsulating complex research applications in Virtual Machines (VMs)
with little or no performance degradation \cite{hepix-vm-benchmark}.
Studies have shown, for example, that high energy physics application code 
runs equally well in a VM or on the native system \cite{chep-vm}.
A key question is how to manage large data sets in a cloud
or distributed cloud environment.   

We have developed a system for running high-throughput batch processing
applications using any number of IaaS clouds \cite{cloud-scheduler}.
The results presented in this work use the IaaS clouds at the 
National Research Council (NRC) in Ottawa, two clouds at the 
University of Victoria (UVIC), and Amazon EC2.
The number of available VM slots at the four clouds was 110 and will be
increased in the near future.  
We also expect to add new clouds to the system in the next few months.
The input data is located on one of the UVIC clouds and the VM repository
is stored on the NRC cloud.
The Canadian sites are connected by a research network provided by CANARIE while the
commodity network is used to connect the NRC and UVIC clouds to Amazon EC2.
To reduce our dependence on the commodity Internet
we made a second copy of the VM repository at Amazon EC2.

We describe the system in the next section, however, we highlight 
the main features.
Users are provided with a set of VMs that are configured
with the application software.  
The user adds code to run their specific analysis 
and saves the VM to a repository.
The user submits their jobs to a scheduler where the job script
contains a link to the required VM.
A component (called Cloud Scheduler) searches the job queue, identifies
the VM required for each queued jobs, and sends out a request to one
of the clouds to boot the user specific VM.
Once the VM is booted, the scheduler submits the user job to the VM.
The jobs runs and returns any output to a user specified location.
If there are no further jobs requiring that specific VM, then
Cloud Scheduler shuts it down.

The system has been demonstrated to work well for applications with 
modest I/O requirements such as the production of 
simulated data \cite{kyle:hpcs}.
The input files for this type of application are small and the rate 
of production of the output data is modest (though the files can be large).
The system has also been used for the analysis of astronomical survey
data where the images are prestaged to the storage
on the cloud \cite{kyle:hpcs}.

In this work, we focus on data intensive high energy physics applications 
where the job reads large sets of input data at higher rates.
We show that the data can be quickly and efficiently
streamed from a single data storage location to each of the clouds.
We will describe the issues that have arisen and the potential for 
scaling the system to many hundreds or thousands of simultaneous user jobs.

\section{System architecture}

\begin{figure*}
\begin{center}
\includegraphics[width=15cm]{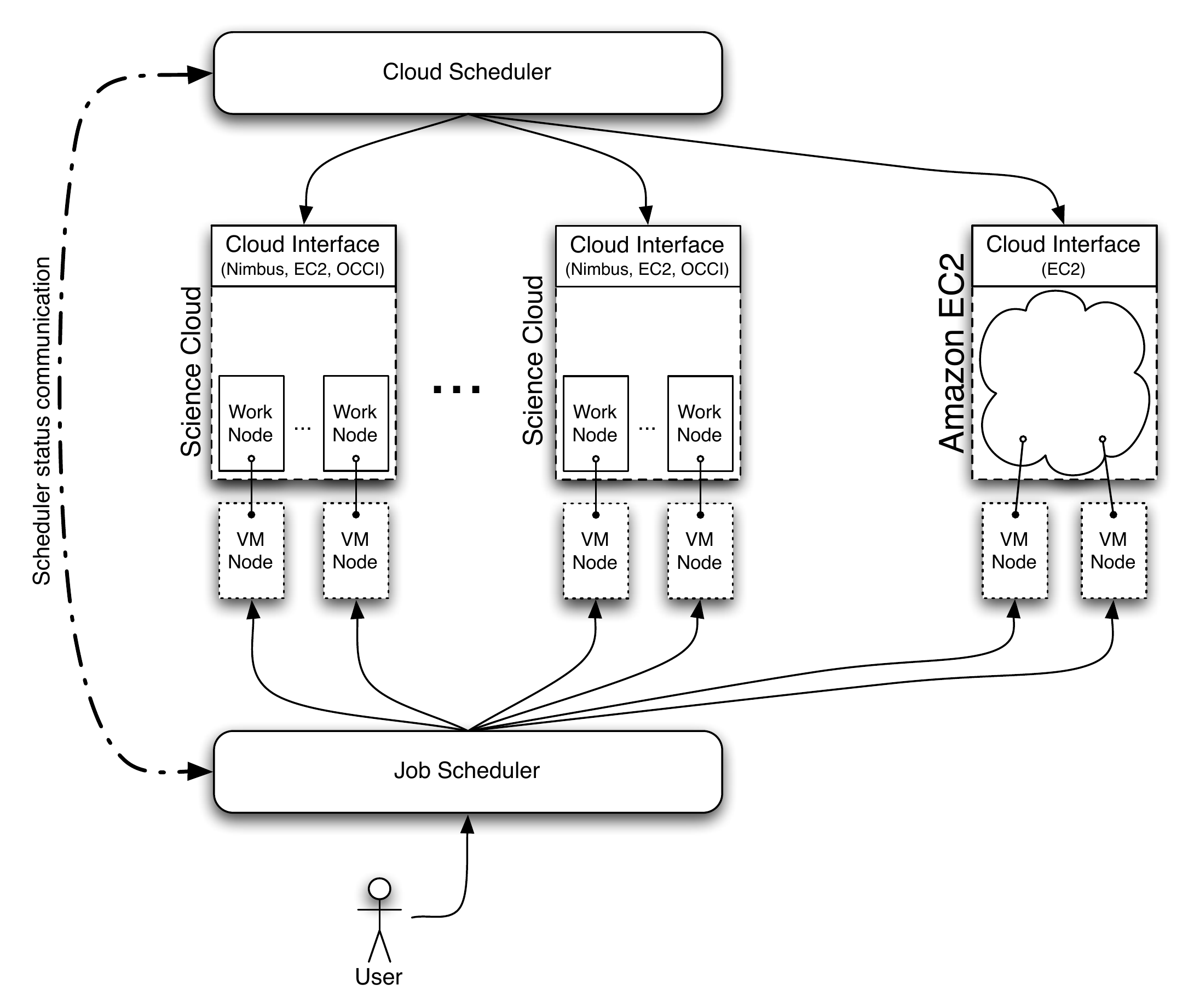}
\end{center}
\caption{\label{fig:overall_arch} 
An overview of the architecture used for the system.
A user prepares their VM image and a job script.  
The job script is submitted to the job scheduler.
The Cloud Scheduler reads the job queue and makes a request to boot the 
user VM on one of the available clouds.
Once there are no more user jobs requiring that VM type, the Cloud 
Scheduler makes a request to the proper cloud to shutdown the user VM.
}
\end{figure*}

The architecture of the distributed IaaS cloud system is shown in 
fig.~\ref{fig:overall_arch} (see ref.~\cite{cloud-scheduler} for a detailed 
description).
The UVIC and NRC clouds use the Nimbus software to manage VMs \cite{nimbus} 
while Amazon EC2 uses its own proprietary software.
The system is designed so that the user submits their jobs to a single scheduler.
The system then boots the user-specific VMs on any one of the available clouds.

Users submit jobs using X.509 Proxy Certificates \cite{rfc3820} to authenticate.  
The certificates are also used to authenticate with Nimbus
clusters when starting, shutting down, or polling VMs. 
Authentication with EC2 is done by using a standard shared access key and 
secret key.

Currently the application VMs are created manually; however, we are
developing a system to simplify the management of user VMs.     
The goal is to have a VM repository where the user can retrieve copies 
of a variety of standard VM images preloaded with the application software.
The user clones the standard VM, adds and tests their software,
saves it in the VM repository and includes a reference to their 
VM image in their job submission script.

The user submits their jobs to the Condor job scheduler \cite{condor}.
We selected Condor as the job schedule since it was designed to utilize
heterogeneous idle workstations and is an ideal job scheduler for a dynamic 
VM environment where workers nodes become available on demand.
Users submit jobs by issuing the \texttt{condor\_submit} command.
The user must add a number of additional parameters specifying the
location of the image and the properties of the required VM.

The management of the VMs on the clouds is done by Cloud Scheduler
\cite{cloud-scheduler}.
Cloud Scheduler monitors the queue of jobs and if one of the jobs 
needs a VM that is not already booted on one of the clouds, then it 
sends a request to the next available cloud.
Cloud Scheduler will also shut down a VM if there are no jobs
needing that particular type of VM.
A complete description of Cloud Scheduler can be found in 
ref. \cite{cloud-scheduler}.

\section{Database and data management}

Analysis jobs in high energy physics typically require two inputs:
event data and configuration data.
The event data can be the real data recorded by the detector or
simulated data.
Each event contains information about the particles seen in detector 
such as their trajectories and energies.
The real and simulated data are nearly identical in format; the simulated 
data contains additional information describing how it was generated.
The user analysis code analyzes one event at a time.
In the BaBar experiment the total size of the real and simulated  
data is approximately 2 PB but users typically read a small
fraction of this sample.
In this work we use a subset of the data containing approximately
2 TB of simulated data and 1.2 TB of real data.  
The configuration data that describes the state of the detector totals 24 GB.

The event data for this analysis was stored in an 8.7 TB 
distributed Lustre file system at UVIC.  
The Lustre file system is hosted on a cluster of six nodes, consisting of a 
Management/Metadata server (MGS/MDS), and five Object Storage servers (OSS).
Each node has dual quad core Intel Nehalem processors and 24 GB of memory.
All servers have six disk drives: two 450 GB, 15K RPM, Serial-Attached 
SCSI (SAS) system drives in a RAID 1 configuration;
and four metadata/object storage drives in a RAID 5 configuration.
The drives for the metadata storage on the MGS/MDS are 450 GB, 15K RPM, 
SAS drives.
The drives for the object storage on each OSS are 1 TB, 
7200 RPM, Serial ATA (SATA) drives. 
Originally configured to allow remote access via mounting over the 
public network, the Lustre filesystem uses a single gigabit interface/VLAN 
to communicate both internally and externally.
This is an important consideration for the test results presented, 
because these same nodes also host the IaaS front-end (MGS/MDT server) 
and Virtual Machine Monitors (OSS servers) for the the UVIC cloud 
and are using the same VLAN to communicate.

The jobs use Xrootd \cite{xrootd} to read the data. 
Xrootd is a file server providing byte level access and is 
used by many high energy physics experiments. 
Xrootd provides read only access to the Lustre data (read/write access is 
also possible) and the capability of implementing GSI authentication. 
Though the implementation of Xrootd is fairly trivial, some optimization was 
necessary to achieve good performance across the network: a read-ahead value 
of 1 MB and a read-ahead cache size of 10 MB was set on each Xrootd client.   

The VM images are stored in a repository hosted on a \emph{lighttpd} file server at NRC.
The VM image is propagated to the worker nodes by http.
As mentioned, we store another copy of the VM images on Amazon EC2 
in an Amazon S3 bucket using the \texttt{ec2\_bundle\_vol} and 
\texttt{ec2\_upload\_bundle} tools provided by the Amazon EC2 API Tools package.

\begin{figure*}[ht]
\begin{center}
\includegraphics[width=15cm]{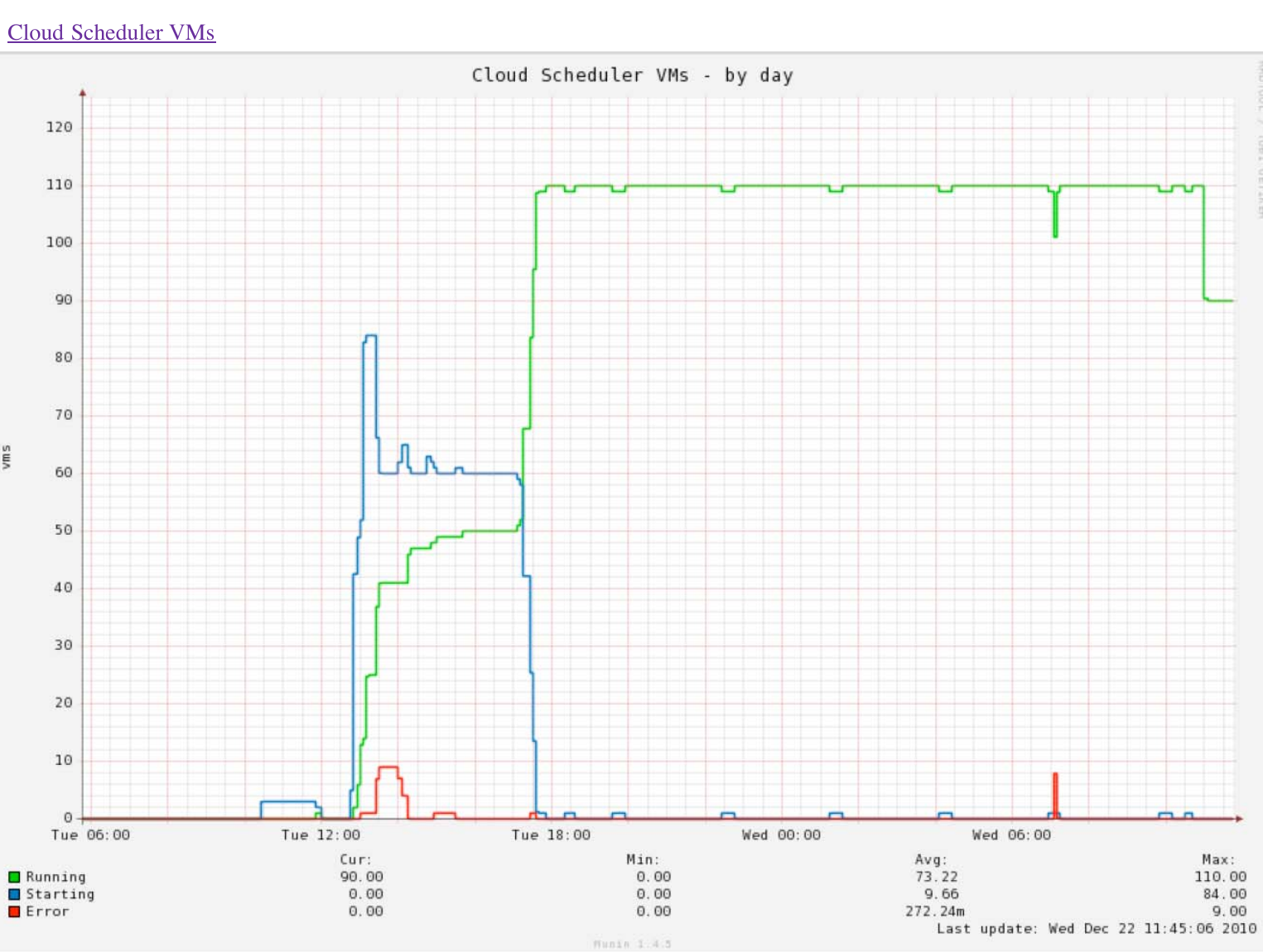}
\end{center}
\caption{\label{fig:vms} 
The plot shows the number of VM's being started (blue line), running 
(green line) and in an error state (red line).
Approximately 250 analysis jobs were submitted on Tuesday at 13:00.
The VMs are located at the NRC cluster and also on Amazon EC2. 
At these sites the VMs booted quickly as can be seen by the green curve.
The two clusters at UVIC needed to retrieve the VMs from NRC in Ottawa 
(3000 km away).
It took approximately 5 hours to transfer the 60 VMs (approximately 1 TB).
A total of 110 VM slots were available.
There were a number of minor errors and they are discussed in the main text.
}
\end{figure*}

\section{Results}

A Xen VM image based on Scientific Linux SL release 5.5 (Boron) was 
created and the latest release of BaBar software was installed.
The BaBar software, which contains C++ and FORTRAN code, requires 1 GB of RAM
and is completely contained in a 16 GB VM. 
In this study, a member of the team and also a researcher in the BaBar 
Collaboration added one of their analysis codes to this VM.
After compiling, linking and testing the code, the modified VM was 
saved for access by the batch system.
A separate VM operated as a head node; it contained the Condor job 
scheduler and Cloud Scheduler for the tests presented in this paper.

A typical user job in high energy physics reads one event at a time 
where the event contains the information of a single particle collision.
Electrons and positrons circulate in opposite directions in a 
storage ring and are made to collide millions of times per second
in the center of the BaBar detector.
The BaBar detector is a cylindrical detector with a size of approximately
5 meters in each dimension. 
The detector measures the trajectories of charged particles and 
the energy of both neutral and charged particles.
A fraction of those events are considering interesting from a scientific
standpoint and the information in the detector is written to a storage medium.
The size of the events in BaBar are a few kilobytes depending on
the number of particles produced in the collision.

The user code analyzes each event independently and writes out a subset 
of the information into an output file for subsequent analysis.
A typical job reads millions of events and the user
submits tens to hundreds of these jobs in parallel.
The output files are a few percent of the original input data set
size, making it easier to study the events of particular interest to the 
researcher.

In this work, the analysis is focused on a particular subset of the BaBar data
set where there are a small number of particles produced in each collisions.
We use three separate samples for our tests which we label Tau1N-data,
Tau1N-MC and Tau11-MC.
The Tau1N-data and Tau1N-MC samples are identical except that the first
is real data and the second is simulated data with a size of 1158 GB
and 615 GB, respectively (the size of each event is 4 KB).
The Tau11-MC sample is a simulated sample with size of 1386 GB
(the size of an event is 3 KB).

We submitted 77 Tau1N-data, 64 Tau1N-MC and 114 Tau11-MC jobs where
a single job runs for approximately 12 hours (the submission of each set
was staggered over a number of hours). 
The number of events processed per second was observed to be 110, 
55 and 430 for the Tau1N-data, Tau1N-MC and Tau11-MC samples, respectively.
This corresponds to an I/O rate of approximately 440, 220 and 
1300 KB/s per job for the three samples.
The processing time is different for each sample as the fraction of interesting
events in each sample is different.
For example, background or non-interesting events are quickly identified and 
discarded so that the job does not waste processing time.

\begin{figure*}[ht]
\begin{center}
\includegraphics[width=16cm]{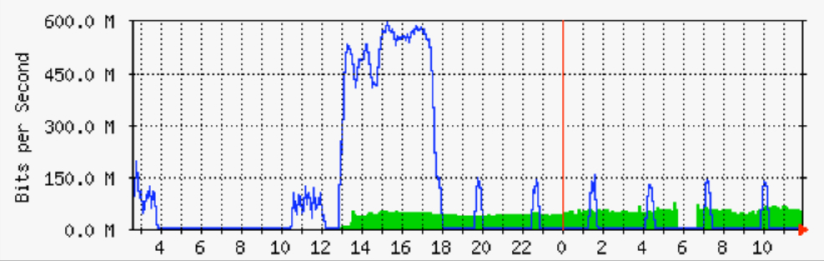}
\end{center}
\caption{\label{fig:nrc}
The inbound and outbound network traffic at the NRC cluster in Ottawa.
The outbound traffic (shown by the blue line) shows the transfer of the 60 VMs
to UVIC (the VMs were stored at NRC and Amazon EC2 has a local copy).
The green histogram shows the data being streamed from UVIC to the 30 jobs 
running on the NRC cluster.
The gap in the green histogram at 06:00 is due to an error in the monitoring 
application.
The periodic spikes in the outbound traffic are due to a VM at UVIC that
failed on a regular basis.
}
\end{figure*}

Fig.~\ref{fig:vms} shows the number of VMs on the system over a 30 hour period.
The green curve is the number of running VMs, the blue curve
is the number of VMs being started or booted and the red curve is the number
of VMs in an error state.
The available number of VM slots were 30 slots at NRC Ottawa, 20 slots
at Amazon EC2, and (40, 20) slots at the two University of Victoria clusters.
The VMs on the NRC and Amazon EC2 clouds started quickly (as they are stored 
locally) whereas the VMs required at UVIC had to be transferred from NRC.
The 60 VMs required at UVIC corresponds to nearly 1 TB and took approximately 
5 hours to transfer (we are working on a method so that only a single copy 
of a VM needs to be transferred).
Fig.~\ref{fig:nrc} shows the network traffic at the NRC cloud.
The VMs required at UVIC were obtained from NRC and the blue curve shows
the outbound network traffic of approximately 500 mbits/s over the
5 hour period (the NRC site has a 1G connection to the CANARIE research 
network in Canada).
Once all the VMs were booted, the outbound network traffic at NRC fell to zero.

\begin{figure*}[ht]
\begin{center}
\includegraphics[width=16cm]{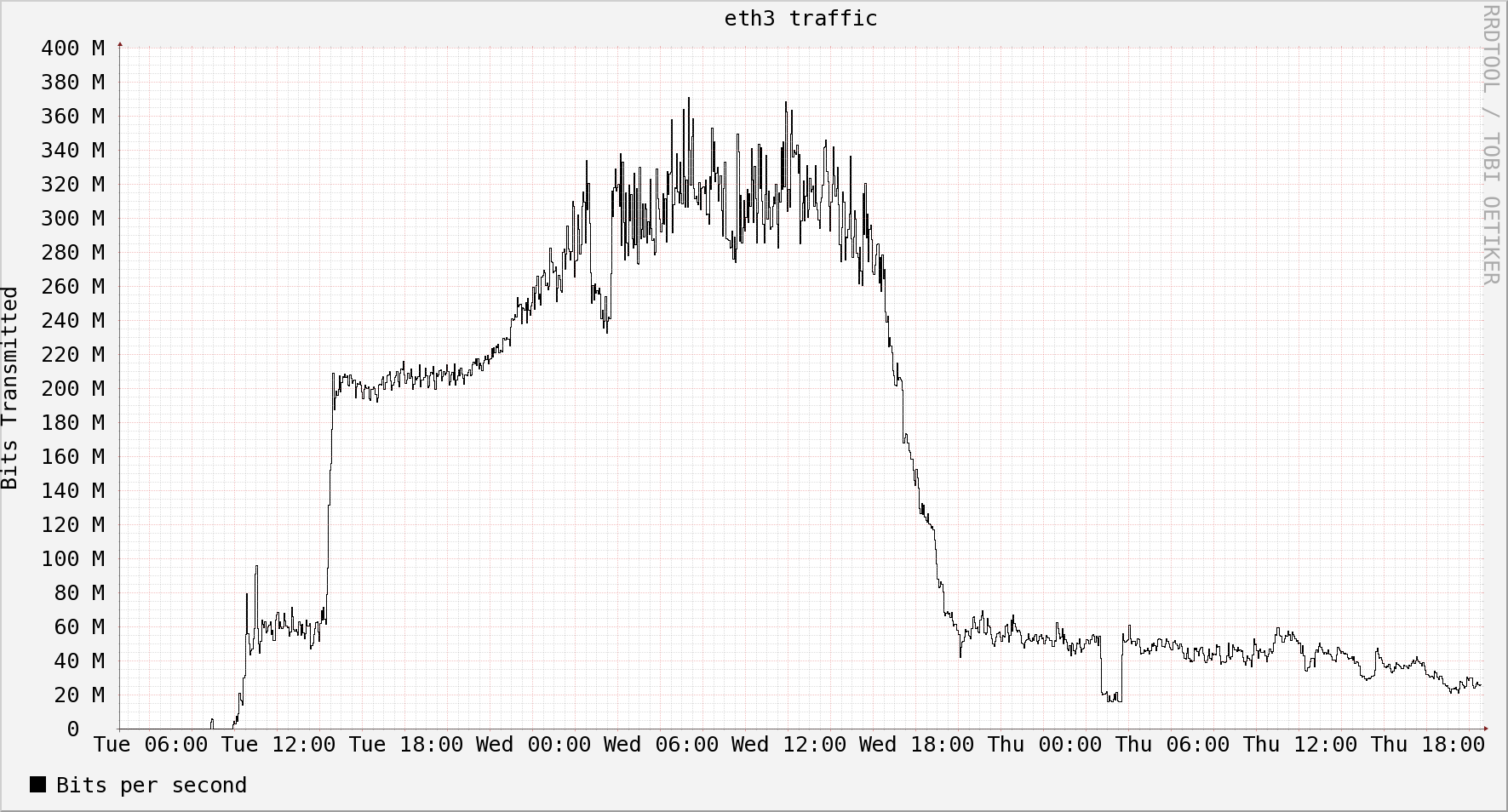}
\end{center}
\caption{\label{fig:elephant}
The outbound network traffic from the Lustre filesystem at UVIC which 
is used to store the event data.
The plot shows the transfer rate to all cloud sites (including the local
UVIC clouds).
The jobs were submitted over a few hour period so that the 
network traffic increases in steps.
The first set of jobs had a lower I/O rate than those submitted later 
so we observe that the network traffic increases with time.
The peak rate is around 320-340 mbits/s for 110 jobs.
}
\end{figure*}

Fig.~\ref{fig:elephant} shows the network bandwidth for the outbound traffic
from the UVIC cluster that hosts all the data.  
The total bandwidth plateaus around 330 mbits/s for the 110 jobs.
Note that the data rate increases over time as the jobs to process the Tau11-MC
sample (which processes the events at the fastest rate) were submitted last and
only started once the first jobs completed. 
The green histogram in fig.~\ref{fig:nrc} shows the inbound data transfer
at 40-50 mbits/s for the 30 jobs at the NRC cloud (note that the gap at 06:00 is
due to error in the monitoring system).

Within a given sample, we find that the time to process the data is 
approximately equal on each of the four clouds.
The UVIC cloud serving the data to the other clouds processes the data at 
the slowest rate.
The other UVIC cloud processes data about 5\% faster than NRC and Amazon EC2
(which are thousands of kilometers from the data source) suggesting that  
the there is little or no latency issues.

One of the features of the system is its ability to recover from faults 
arising from local system problems at each of the clouds.   
We list some of the problems we identified in the processing of the jobs.
For example, we find that:

\begin{itemize}
\item
A number of VMs at NRC went into an error state 
(see the red line in fig.~\ref{fig:vms} at Tuesday 13:00) and
this was attributed to a bug in the older version of Nimbus used at NRC that
has been fixed in the latest release.
In any case, Cloud Scheduler killed those VMs and restarted new ones.
The red spike at Wednesday 07:00 in fig.~\ref{fig:vms}  was due to a loss of 
communication between the VMs on one of the clouds and Cloud Scheduler;
once the link was restored the correct status was obtained.

\item
A VM on one of the UVIC clouds failed on a regular basis (this can 
be seen if fig.~\ref{fig:nrc} where there are spikes in the outbound
NRC traffic as the VM  was resent).
Messages in the Xen daemon log indicate that the failure of this VM
was initiated by a \texttt{shutdown} command followed ten seconds later by a 
\texttt{destroy} command issued to the Xen hypervisor.
The source of these commands is not identified.
Future tests will incorporate additional diagnostics to identify and eliminate this error.

\item
At the right end of fig.~\ref{fig:vms}, the number of VMs dropped
from 110 to 90 as we stopped the jobs on the Amazon EC2 cluster.
We noted a problem with the configuration of Condor for EC2;  it was 
corrected and reactivated (the restart of the EC2 VMs is not shown
in fig.~\ref{fig:vms}).
\end{itemize}

The overall system has preformed remarkably well with all jobs 
processed successfully.
We are planning to upgrade the network of the data storage system to 
10 gbits/s and this would allows us to scale the number of jobs 
by an order of magnitude.
There are also plans in 2011 to connect Amazon EC2 to the Canadian research
networked (provided by CANARIE).

A significant fraction of the data transfer in this test was the 
movement of VMs.  
We will soon implement a system so that only a single copy of the VM
image would need to be transferred, significantly reducing the start-up
time of the remote sites.

\section{Conclusion}

We have presented the results for running a data intensive
particle physics applications in a distributed compute cloud.
The results show that 4-5 TB data sets can be easily accessed
from remote locations over a high-speed research network.
There appear to be no limitations to scaling the number of jobs
by an order of magnitude once we increase the network bandwidth
from the event storage servers.
From the users perspective, the system is robust and the success rate
of job completion is close to 100\% in tests to date.
We have shown that the use of distributed compute clouds can be an effective way of 
analyzing large research data sets.

\section*{Acknowledgment}
The support of CANARIE, the Natural Sciences and Engineering Research Council,
the National Research Council of Canada and Amazon are acknowledged.



\begin{thebibliography}{1}

\bibitem{hepix-vm-benchmark}
M. Alef and I. Gable.
HEP specific benchmarks of virtual machines on multi-core CPU architectures.
J. Phys Conf. Ser. 219, 052015 (2008).  
doi: 10.1088/1742-6596/219/5/052015

\bibitem{chep-vm}
A. Agarwal, A. Charbonneau, R. Desmarais, R. Enge, I. Gable, D. Grundy, 
D. Penfold-Brown, R. Seuster, R.J. Sobie, and D.C. Vanderster.
Deploying HEP Applications Using Xen and Globus Virtual Workspaces. 
J. Phys.: Conf. Ser. 119, 062002 (2008).
 doi: 10.1088/1742-6596/119/6/062002

\bibitem{cloud-scheduler}
P.Armstrong, A.Agarwal, A.Bishop, A.Charbonneau, R.Desmarais, K.Fransham, N. Hill, 
I.Gable, S.Gaudet, S.Goliath, R.Impey, C.Leavett-Brown, J.Ouellete, M.Paterson, 
C.Pritchet, D.Penfold-Brown, W.Podaima, D.Schade, R.J.Sobie. 
Cloud Scheduler: a resource manager for a distributed compute cloud, 
June 2010 ( arXiv:1007.0050v1 [cs.DC] )

\bibitem{kyle:hpcs}
K.Fransham, P.Armstrong, A.Agarwal, A.Bishop, A.Charbonneau, R.Desmarais, N. Hill, 
I.Gable, S.Gaudet, S.Goliath, R.Impey, C.Leavett-Brown, J.Ouellete, M.Paterson, 
C.Pritchet, D.Penfold-Brown, W.Podaima, D.Schade, R.J.Sobie. 
Research computing in a distributed compute cloud, 
Proceedings of the High Performance Computing Symposium, Toronto 2010 
(doi: 10.1088/1742-6596/256/1/012003)

\bibitem{nimbus}
K.Keahey, I.Foster, T.Freeman, and X. Zhang.
Virtual workspaces: Achieving quality of service and quality of life in the Grid. 
Sci. Program. Vol. 13 (2005) 265.

\bibitem{rfc3820}
S.Tuecke, V.Welch, D.Engert, L.Pearlman,and M.Thompson.
Internet X.509 Public Key Infrastructure (PKI) Proxy Certificate Profile.
http://www.ietf.org/rfc/rfc3820.txt.

\bibitem{condor}
D.Thani, T.Tannenbaum and M.Livny.
Distributed computing in practice: the Condor experience.
Concurrency and Computation: Practice and Experience
Vol. 17 (2005) 323.

\bibitem{xrootd}
Alvise Dorigo, Peter Elmer, Fabrizio Furano, and Andrew Hanushevsky. 2005. 
XROOTD/TXNetFile: a highly scalable architecture for data access in the ROOT environment. 
In Proceedings of the 4th WSEAS International Conference on Telecommunications 
and Informatics (TELE-INFO'05)

\end{thebibliography}
\end{document}